\documentclass[aps,pra,twocolumn,showpacs,superscriptaddress]{revtex4-1}

\usepackage{amsfonts,mathrsfs,amsmath,amsthm,amssymb}
\usepackage{bm,times}
\usepackage{graphicx,epsfig}
\usepackage[colorlinks=true,breaklinks=true,linkcolor=blue,citecolor=blue,urlcolor=blue]{hyperref}

\newcommand{\be}{\begin{eqnarray}}
\newcommand{\ee}{\end{eqnarray}}

\makeatletter
    
    \newcommand{\Rmnum}[1]{\expandafter\@slowromancap\romannumeral #1@}
\makeatother

\begin{document}

\author{Xiaohua Wu}
\email{wxhscu@scu.edu.cn}
\affiliation{College of Physical Science and Technology, Sichuan University, Chengdu 610064, China}

\author{Tao Zhou}
\email{taozhou@swjtu.edu.cn}
\affiliation{School of Physical Science and Technology, Southwest Jiaotong University, Chengdu 611756, China}

\title{Diagnosing steerability of a bipartite state with the nonsteering threshold}

\date{\today}

\begin{abstract}
In the present work, a traditional quantity, the averaged fidelity, is introduced as the steering parameter. From the definitions of steering
from Alice to Bob and the joint measurability, a general scheme is developed to design linear steering criteria. For a given set of measurements on Bob's side, the so-called nonsteering threshold is defined to quantify the ability to detect steering. If the measured averaged fidelity exceeds this threshold, it is shown that the state shared by Alice and Bob is steerable from Alice to Bob, and the measurements performed by Alice are also verified to be incompatible. Within the general scheme, a discussion about how to design optimal criteria is provided for two different scenarios: (a) to find the optimal measurements on Bob's side when the state is unknown and (b) to find the optimal measurements for Alice when the state and the measurements on Bob's side are given.
\end{abstract}

\maketitle

\section{introduction}
The  concept of \emph{steering} was introduced by Schr\"odinger in 1935~\cite{Sch}  as a generalization of the Einstein-Podolsky-Rosen (EPR) paradox~\cite{Ein}. Steering infers the fact, in a bipartite scenario, an observer can effect the state of a far remote system through local measurement. Specifically, if Alice and Bob share an entangled state, Alice can remotely steer Bob's state by performing measurements only in her part of the system. In 2007, Wiseman, Jones, and Doherty~\cite{Wiseman1} formally defined  quantum steering as a type of quantum nonlocality that is logically different from nonseparability~\cite {Guhne,Horos} and Bell nonlocality~\cite{Brunner}. Quantum steering can be understood as the failure of a hybrid \emph{local-hidden-variable} (LHV)--\emph{local hidden state} (LHS) model to simulate quantum correlations between Alice and Bob.

Unlike quantum nonlocality and entanglement,  EPR steering is inherently asymmetric with respect to the observers. It is shown that there exist entangled states which are one-way steerable~\cite{bowles,Bow}. Besides its foundational significance, EPR steering has a vast range of information-theoretic applications in one-sided device-independent scenarios, where the party being steered has trust on his or her own quantum devices while the other's devices may be untrusted. These applications range from one-sided device-independent quantum key distribution~\cite{Bran}, advantage in subchannel discrimination~\cite{piani}, secure quantum teleportation~\cite{Reid1,He}, quantum communication~\cite{Reid1}, detecting bound-entanglement~\cite{Mor}, and one-sided device-independent randomness generation~\cite{law}, to one-sided device-independent self-testing of pure maximally as well as nonmaximally entangled states~\cite{supic}.

In 1989, Reid derived variance inequalities that are violated with EPR correlations for a continuous variable system~\cite{eid} and this was extended to discrete variable systems in Ref.~\cite{Caval}. In quantum information~\cite{Wiseman1}, EPR steering can be defined as the task for a referee to determine whether two parties share entanglement. Based on these, EPR-steering inequalities were defined in Ref.~\cite{can22}, with the property that violation of any such inequality implies steering. After this  work, further tools have been proposed to signalize steering from experimental correlations, for instance, the linear and nonlinear steering criteria~\cite{sau,wit,Evan,mar,rut}, steering criteria from uncertainty relations~\cite{wa,schnee,Costaa,costab,jia,kri}, steering with
 Clauser-Horne-Shimony-Holt (CHSH)-like inequalities~\cite{Can3,Girdhar,cos,quan}, moment matrix approach~\cite{Kig,mo,chen00}, and steering criteria based on local uncertainty relations~\cite{Ji,Zhen}. The discussed criteria or small variation thereof have been used in several experiments~\cite{sau,wit,Bennet,smith,weston} .

Joint  measurability was extensively studied for a few decades before steering was formulated in its modern form. Operationally, it corresponds to the possibility of deducing the statistics of several measurements from the statistics of a single one. In recent years, it has been shown that steering is related to joint measurability in a many-to-one manner~\cite{quint,ula,UULA,Kiukas}. The connection between the joint measurability and steering unlocks the technical machinery developed within the framework of quantum measurement theory used in the context of quantum correlations~\cite{rmd}.

Our  work is originated from one of the open problems summarized in~\cite{rmd}: In experiments, quantum steering is verified by a finite number of measurement settings. There are only a few works on optimizing the measurement settings (when having a fixed number of inputs or outputs), and more research is needed to serve as input for experiments. Here, we shall focus on the case where the linear steering
inequalities are applied for detecting steering~\cite{can22, sau,Joness}.

In the present work, we introduce a traditional quantity, the averaged fidelity, as the steering parameter. From the definitions of steering from Alice to Bob~\cite{Can1} and the joint measurability, we propose a general scheme to design the linear steering criteria. For a given set of
measurements on Bob's side, we define a quantity, the so-called non-steering-threshold (NST), to quantify its ability to detect steering. If the measured averaged fidelity exceeds this threshold, the state shared by Alice and Bob is steerable from Alice to Bob, and the measurements performed by Alice are also verified to be incompatible. Within the general scheme, we give a discussion about how to design optimal criteria for two different scenarios as follows. (a) When the state is unknown, which measurement is optimal on Bob's side? (b) When the state and the
measurements on Bob's side are given, which measurement is optimal for Alice?

The content of the present work is organized as follows. In Sec.~\ref{Sec2}, we give a brief review of the definitions of steering from Alice to Bob and the joint measurability. In Sec.~\ref{Sec3}, details about the nonsteering threshold are discussed. In Sec.~\ref{Sec4}, we address the problem of constructing optimal linear criteria for the case that the state is unknown. The optimal set of measurements on Alice's side is discussed in Sec.~\ref{Sec5}.  Finally, we end our work with a short conclusion.

\section{Preliminary}
\label{Sec2}
\subsection{Steering from Alice to Bob}

Before showing how to apply the mentioned protocol to demonstrate whether a state is steerable from Alice to Bob, it is necessary to introduce some denotation at first. For a matrix $C=\sum_{i,j}c_{ij}\vert i \rangle \langle j\vert$, $C^{\mathrm{T}}=\sum_{ij}c_{ij}\vert j\rangle\langle i\vert$ is the transposed matrix, $C^*=\sum_{i,j}c_{ij}^*\vert i\rangle\langle j \vert$ denotes the complex conjugation of $C$, and $C^{\dagger}$, which is defined as $C^{\dagger}=(C^*)^{\mathrm{T}}$, can be calculated as  $C^{\dagger}=\sum_{ij}c_{ij}^*\vert j\rangle\langle i\vert.$

In the field of quantum information and quantum computation, entanglement is one of the most important quantum resources. To verify whether a bipartite state $W$ is entangled, it is usually expressed with a pure entangled state and a one-sided quantum channel~\cite{horo1,Ruskai},
\begin{equation}
W=\mathbb{I}_d\otimes \varepsilon(\vert \Psi\rangle\langle \Psi\vert),
\end{equation}
where $\mathbb{I}_d$ is an identity map, and $\vert\Psi\rangle=\sum _{i=1}^d \sqrt{\lambda_i}\vert i\rangle\otimes \vert i\rangle$ is the purification of the reduced density matrix for Alice, say, $\rho_{\mathrm{A}}=\sum_{i=1}^d\lambda_i\vert i\rangle\langle i\vert$ with $d$ the dimension of the Hilbert space $\mathcal{H}$. Based on this decomposition, the state $W$ should be a mixture of product states if and only if the channel $\varepsilon$ is entanglement-breaking (EB)~\cite{horo1,Ruskai}.

Consider an unknown quantum state $W$ shared by Alice and Bob. Alice can perform $N$ measurements on her subsystem labeled by $\mu=1,2,...,N$, each having $d$ outcomes $a=1,2,...,d$. The $\mu$-th projective measurement is denoted by $\hat{\Pi}^{a}_{\mu}$, and $\sum_{a=1}^{d}\hat{\Pi}^{a}_{\mu}=I_d $, with $I_d$ the identity operator for a $d$-dimensional Hilbert space. Now, $\hat{\Psi}^{a}_{\mu}$, usually referred to as the input state, can be defined from the diagonal density matrix $\tau$ and the $\hat{\Pi}^{a}_{\mu}$,
\begin{equation}
\hat{\Psi}^{a}_{\mu}=\frac{\sqrt{\rho_\mathrm{A}}(\Pi_{\mu}^a)^*\sqrt{\rho_\mathrm{A}}}{\mathrm{Tr}(\hat{\Pi}^{a}_{\mu}\rho_\mathrm{A})},
\end{equation}
with the probability $p(a\vert\mu)=\mathrm{Tr}(\hat{\Pi}^{a}_{\mu}\rho_\mathrm{A})$ and $\sum_{a=1}^d p(a\vert\mu)=1$. Upon performing $\mu$-measurement and receiving outcome $a$, the state of Bob's subsystem becomes the state $\rho_{\mu}^{a}$ with probability $p(a\vert\mu)$, where the normalized state $\rho^{a}_{\mu}=\varepsilon(\hat{\Psi}^{a}_{\mu})$ can be regarded as the output of the channel $\varepsilon$ with $\hat{\Psi}^{a}_{\mu}$ as the input. Now, the set of unnormalized quantum states $\{\tilde{\rho}^{a}_{\mu}\}_{a,\mu}$ with $\tilde{\rho}^{a}_{\mu}=p(a\vert\mu)\rho_{\mu}^{a}$ is usually called an \emph{assemblage}.

In 2007, Wiseman, Jones, and Doherty \cite{Wiseman1} formally defined quantum steering as the possibility of remotely generating ensembles that could not be produced by a \emph{local~hidden~state} (LHS) model.  A LHS model refers to the case where a source sends a classical message $\xi$ to one of the parties, say, Alice, and a corresponding quantum state $\rho_{\xi}$ to another party, say Bob, and when Alice performs the  $\mu$-th measurement, the variable $\xi$ instructs the output $a$ of Alice's apparatus with the probability $\mathfrak{p}(a\vert\mu,\xi)$. In addition, it is usually considered that the variable $\xi$ is chosen according to a probability distribution $\omega(\xi)$ with $\int\omega(\xi)d\xi=1$. Bob does not have access to the classical
variable $\xi$, his final assemblage is composed by
\begin{equation}
\label{assemblage}
\tilde{\rho}^{a}_{\mu}=\int\omega(\xi)d\xi\mathfrak{p}(a\vert\mu,\xi)\rho_{\xi},
\end{equation}
and $p(a\vert\mu)$, the probability that the outcome is $a$ when a measurement $\mu$ is performed by Alice, is expressed as $p(a\vert\mu)=\int\omega(\xi)d\xi\mathfrak{p}(a\vert\mu,\xi)$.
The definition of steering in the present work comes directly from the review article \cite{Can1}: an assemblage is said to demonstrate steering if it does not admit a decomposition of the form in Eq.~\eqref{assemblage}.  Furthermore, a quantum state $W$ is said to be steerable from Alice to Bob if the experiments in Alice's part produce an assemblage that demonstrate steering. On the contrary,\emph{ an assemblage is said to be LHS if it can be expressed as in Eq.~\eqref{assemblage}, and a quantum state is said to be unsteerable if a LHS assemblage is generated for all local measurements.}
\subsection{Critical visibility}
As it was shown in Refs.~\cite{quint,ula,UULA,Kiukas}, steering is related to incompatibility  in a many-to-one manner. A set of measurements $\{\hat{M}_{\mu}^a\}$, where
$\mu\in\{1,2,...,N\}$ labels the measurements in the set and $a\in\{1,...,k\}$ labels the outcomes of each measurement, is jointly measurable,
or compatible, if there exists a joint measurement $\{\hat{M}_\lambda\}$, such that
\begin{equation}
\hat{M}_{\mu}^a=\sum_{\lambda}\pi(\lambda)p(a\vert\mu,\lambda)\hat{M}_\lambda,
\end{equation}
for all $a$ and $\mu$, with $\pi(\lambda)$ and $p(a\vert\mu,\lambda)$ the probability distributions. From it, all positive-operator-valued measurements (POVMs) $\hat{M}_{\mu}^a$ can be recovered by coarse-graining over the joint measurement $\{\hat{M}_\lambda\}$.

By introducing critical visibility, a quantity used to characterize the white-noise robustness of an assemblage, Bavaresco \textit{et.~al.}~\cite{bava} recently addressed the problem of determining whether a quantum state can demonstrate steering when subjected to $N$ measurements with $k$ outcomes. The general methods have been proposed to calculate the upper bounds for the white-noise robustness of a $d$-dimensional quantum state under different measurement scenarios. Here, we shall focus on one of the discussed cases where the assemblage is given by the  isotopic state with projective measurements. Considering the depolarizing map $\varepsilon_{\eta}$ acting on a Hermitian operator $\hat{A}$ of a $d$-dimensional Hilbert space $\mathcal{H}$
\begin{equation}
\varepsilon_{\eta}: \hat{A}\rightarrow\varepsilon_{\eta}(\hat{A})=\eta\hat{A}+(1-\eta)\mathrm{Tr}(\hat{A})\frac{I_d}{d},
\end{equation}
and with $\vert\phi^{+}_d\rangle=\frac{1}{\sqrt{d}}\sum_{i=1}^d\vert ii\rangle$ the maximally entangled state, the isotopic state $W_{\eta}$ in $\mathcal{H}\otimes\mathcal{H}$ can be defined as
\begin{equation}
\label{isotopic}
W_{\eta}:= \mathbb{I}_d\otimes\varepsilon_{\eta}(\vert\phi^+_d\rangle\langle\phi^+_d\vert)=\eta\vert\phi^+_d\rangle\langle\phi^+_d\vert+(1-\eta)\frac{I_{d^2}}{d^2}.
\end{equation}
For $d=2$, it coincides with the two-qubit Werner state~\cite{Werner}. Assume that Alice performs the projective measurements $\{\hat{M}_{\mu}^{a*}\}$, with $\mu=1,2,..,N$ and $a=1,2,...,d$, and the resulted assemblage $\{\tilde{\rho}^{a}_{\mu}\}$ is
\begin{equation}
\tilde{\rho}^{a}_{\mu}=\frac{1}{d}\varepsilon_{\eta}(\hat{M}_{\mu}^{a}).
\end{equation}
The critical visibility $\eta(\{\hat{M}_{\mu}^{a}\})$ is defined as
\begin{equation}
\label{CV}
\eta(\{\hat{M}_{\mu}^{a}\}):=\max\left\{\eta \vert \{\varepsilon_{\eta}(\hat{M}_{\mu}^{a})\}_{a,\mu}\in \mathrm{LHS}\right\}.
\end{equation}
where $\mathrm{LHS }$ is the set of assemblages that admit a LHS model. Therefore, $\eta(\{\hat{M}_{\mu}^{a}\})$ is the \emph{exact value} above
 which $\{\varepsilon_{\eta}(\hat{M}_{\mu}^{a})\}$ no longer admits an LHS model. From the definition above, the critical visibility can also be viewed as a quantifier for the ability of the set of measurements $\{\hat{M}_{\mu}^{a}\}$ to go against the white noise~\cite{bava}.

With the quantity $\eta(\{\hat{M}_{\mu}^{a}\})$ introduced above, one may ask the following. When the number of the experiment settings and the outcomes are fixed, which measurement is the most incompatible? This question is equivalent to calculating the best critical visibility $\eta^*(\{\hat{M}_{\mu}^{a}\}^{{\min}}, N,d)$ over all the possible projective measurements,
\begin{equation}
\eta^*(\{\hat{M}_{\mu}^{a}\}^{\mathrm{\min}}, N,d):=\min_{\{\hat{M}_{\mu}^{a}\}}\eta(\{\hat{M}_{\mu}^{a}\}),
\end{equation}
with $\{\hat{M}_{\mu}^{a}\}^{\mathrm{\min}}$ being the most incompatible set  of the  measurements. The calculation of $\eta^*(\{\hat{M}_{\mu}^{a}\}^{\mathrm{\min}}, N,d)$ is a difficult task, and only for a few cases, have the exact values been obtained~\cite{bava}.

 \section{Non-steering threshold}
 \label{Sec3}
\subsection{ Sufficient criteria for steering}
For a given output $\rho^{a}_{\mu}$ , we suppose that Bob will measure the fidelity with a set of rank-one projective operators $\{\hat{M}^a_{\mu}\}$,
\begin{equation}
\hat{M}^a_{\mu}:=\hat{\Phi}^a_{\mu}=\vert \phi^a_{\mu}\rangle\langle\phi^a_{\mu}\vert,\ \langle\phi^a_{\mu}\vert\phi^a_{\mu}\rangle=\delta_{ab},\ \sum_{a=1}^d\hat{M}_{\mu}^{a}=I_d,
\end{equation}
which  are usually called target states in previous works. The fidelity $F(a\vert\mu)$ is defined as the overlap between the target and the un-normalized conditional state, $F(a\vert\mu)=\mathrm{Tr}[\hat{\Phi}^a_{\mu}p(a\vert{\mu})\varepsilon(\hat{\Psi}^a_{\mu})]$.
Let $\langle A\otimes B  \rangle =\mathrm{Tr}[W(A\otimes B)]$ be the expectation value of the operator $A\otimes B$, and, in experiment, $F(a\vert\mu)$ can be measured as
\begin{equation}
F(a\vert\mu)=\langle \hat{\Pi}^a_{\mu}\otimes \hat{\Phi}^a_{\mu}\rangle.
\end{equation}

Denote $q_{\mu}$ the probability that the $\mu$-th measurement has been performed, $\sum_{\mu=1}^N q_{\mu}=1$, and then, the averaged fidelity $F_{\mathrm{avg}}$ can be expressed as
\begin{equation}
F_{\mathrm{avg}}=\sum_{\mu=1}^N\sum_{a=1}^d q_{\mu}F(a\vert\mu).
\end{equation}

For an EB channel $\varepsilon_{\mathrm{EB}}$, similarly to the above definitions of $F(a\vert\mu)$ and $F_{\mathrm{avg}}$, one can introduce $F^{\mathrm{EB}}(a\vert\mu)=\mathrm{Tr}[\hat{\Phi}^a_{\mu}p(a\vert{\mu})\varepsilon_{\mathrm{EB}}(\hat{\Psi}^a_{\mu})]$ and $F^{\mathrm{EB}}_{\mathrm{avg}}=\sum_{\mu}\sum_{a} q_{\mu}F^{\mathrm{EB}}(a\vert\mu)$, and the classic fidelity threshold (CFT) can be defined as
\begin{equation}
\mathfrak{F}_{\mathrm{CFT}}=\max_{\{ \varepsilon_{\mathrm{EB}}\}}F^{\mathrm{EB}}_{\mathrm{avg}},
\end{equation}
where the optimization is taken over the set $\{ \varepsilon_{\mathrm{EB}}\}$ of all EB channels. As shown in previous works~\cite{Barnett1,Fuchs,Massar,Horo,Adesso1,Chir,Namiki3,Chir1,Namiki4}, the CFT  depends on the actual choices of the input and target states. If the experiment result $F_{\mathrm{avg}}$ exceeds this threshold, $F_{\mathrm{avg}}>\mathfrak{F}_{\mathrm{CFT}}$,
one may conclude that the channel does not belong to the set $\{ \varepsilon_{\mathrm{EB}}\}$, and the state $W$ is an entangled state.

For the task of detecting steering, a criterion can be introduced in a similar way. Consider that the assemblage $\{p(a\vert{\mu})\varepsilon(\hat{\Psi}^a_{\mu}\}_{\mu,a}\}$  has an LHS decomposition, and then one can define
\begin{eqnarray}
F^{\mathrm{LHS}}_{\mathrm{avg}}\equiv\int\omega(\xi)\mathrm{Tr}[\rho_{\xi}\bar{\rho}]d\xi,
\end{eqnarray}
where
\begin{eqnarray}
\label{rhobar}
\bar{\rho}=\sum_{\mu=1}^N\sum_{a=1}^{d}q_{\mu}\mathfrak{p}(a\vert\mu,\xi)\hat{\Phi}^{a}_{\mu},
\end{eqnarray}
with the probability $\mathfrak{p}(a\vert\mu,\xi)$ interpreted as the value of $\hat{\Pi}^a_{\mu}$ in the LHV model. For the set of $\mu$-th measurement operators $\{\Pi^{a}_{\mu}|a=1,2,...,d\}$, $\sum_{a=1}^d\Pi^{a}_{\mu}=I_d$,
there should be $\sum_{a=1}^d\mathfrak{p}(a\vert\mu,\xi)=1$. It can be seen that $\bar{\rho}$ in Eq.~(\ref{rhobar}) is a density matrix, and it can be formally decomposed as $\bar{\rho }=\sum_{\nu}\lambda_{\nu}\vert \lambda_\nu\rangle\langle \lambda_{\nu}\vert$, with $\lambda_{\nu}$ the eigenvalues and $\vert \lambda_{\nu}\rangle $ the corresponding eigenvectors. Denote the largest eigenvalue by the cross norm $\bar{\rho}_{\times}=\max_{\vert\phi\rangle}\langle\phi\vert\bar{\rho}\vert\phi\rangle$, and one can define the non-steering threshold (NST) as
\begin{equation}
\mathfrak{F}_{\mathrm{NST}}(\{q_{\mu},M^a_{\mu}\},N, d)=\max_{\{\mathfrak{p}(a\vert\mu,\xi)\}_{a,\mu}}\bar{\rho}_{\times}.
\end{equation}
Here, $\bar{\rho}_{\times}$ is a function of the variables $\mathfrak{p}(a\vert\mu,\xi)$, and $\mathfrak{F}_{\mathrm{NST}}$ is the maximum value of $\bar{\rho}_{\times}$. Together with the result $\mathrm{Tr}[\rho_{\xi}\bar{\rho}]\leq\mathfrak{F}_{\mathrm{NST}}$ and $\int \omega(\xi)d\xi=1$, it can be found that $\mathfrak{F}_{\mathrm{NST}}$ is an upper bound of $F^{\mathrm{LHS}}_{\mathrm{avg}}$, say $\mathfrak{F}_{\mathrm{NST}}\geq F^{\mathrm{LHS}}_{\mathrm{avg}}$. Therefore, once the experiment result $F_{\mathrm{avg}}$ exceeds this threshold
\begin{equation}
\label{criteria}
F_{\mathrm{avg}}> \mathfrak{F}_{\mathrm{NST}}(\{q_{\mu},M^a_{\mu}\},N, d),
\end{equation}
the assemblage $\{p(a\vert{\mu})\varepsilon(\hat{\Psi}^a_{\mu})\}_{\mu,a}$ does not admit an LHS decomposition and the state $W$ is steerable from Alice to Bob. For the set of measurements $\{q_{\mu}, M^a_{\mu}\}$,  its ability to detect steering is quantified by $\mathfrak{F}_{\mathrm{NST}}(\{q_{\mu},M^a_{\mu}\},N, d)$. Here, it should be emphasized that the way to calculate $\mathfrak{F}_{\mathrm{NST}}(\{q_{\mu},M^a_{\mu}\},N, d)$ is based on an elementary idea proposed in Refs.~\cite{can22, sau,Joness}, where a steering inequality can be constructed by just considering the measurements performed by Bob. The similar method can also be found in the works on incompatibility breaking quantum channels~\cite{Kiukas,hein}.

According to the work in Ref.~\cite{Can1}, the assemblage $\{ p(a\vert\mu)\varepsilon_{\mathrm{EB}}(\hat{\Psi}^{a}_{\mu})\}_{a,\mu}$ always admits an LHS decomposition and, therefore, there exists a relation
\begin{equation}
\mathfrak{F}_{\mathrm{NST}}(\{q_{\mu},M^a_{\mu}\},N, d)\ge\mathfrak{F}_{\mathrm{CFT}},
\end{equation}
which is nothing but a quantitative expression for the conclusion that steering is stronger than entanglement.

In this work, $\mathfrak{F}_{\mathrm{NST}}$ is usually required to be a tight-bound, since it can always be achieved with an assemblage $\{p(a\vert\mu) \varepsilon(\hat{\Psi}^a_{\mu})\}_{a,\mu}$ admitting a LHS model. This requirement makes the criteria in Eq.~\eqref{criteria} a necessary and sufficient condition for detecting steering. Finally, the so-called geometric averaged fidelity is introduced as
\begin{equation}
f_{\mathrm{avg}}:=\frac{dF_{\mathrm{avg}}-1}{d-1},
\end{equation}
and we fix the value of the parameter $\eta$ with $f_{\mathrm{avg}}$,
 $\eta=f_{\mathrm{avg}}$.
Correspondingly, we define $\mathfrak{g}_{\mathrm{NST}}(\{q_{\mu},M^a_{\mu}\},N, d)$, the so-called geometric NST, as
\begin{equation}
\label{GNST}
\mathfrak{g}_{\mathrm{NST}}(\{q_{\mu},M^a_{\mu}\},N, d)=\frac{d\mathfrak{F}_{\mathrm{NST}}(\{q_{\mu},M^a_{\mu}\},N, d)-1}{d-1},
\end{equation}
and the criterion in Eq.~\eqref{criteria} can be expressed in an equivalent form:
\begin{equation}
\eta>  \mathfrak{g}_{\mathrm{NST}}(\{q_{\mu},M^a_{\mu}\},N, d).
\end{equation}
If this inequality is satisfied, it is obvious that the state $W$ is steerable from Alice to Bob.

\subsection{The relation between the geometric NST and the critical visibility}
Let us firstly consider the case where the state $W$ is unknown while the probability $q_{\mu}$ is fixed as $q_{\mu}=1/N$. The following two statements---(a) the state is steerable from Alice to Bob and (b) the set of measurements $\{\Pi^a_{\mu}\}$ performed by Alice is incompatible---are necessary so that the assemblage  $\{{\tilde{\rho}}^{a}_{\mu}\}$ does not admit a LHS  model. Therefore, the criterion $\eta> \mathfrak{g}_{\mathrm{NST}}(\{M^a_{\mu}\},N, d)$, is a sufficient condition for Bob to declare the statements (a) and (b). There are many possible protocols to obtain the same averaged fidelity, and, as one of them, one may  suppose that the state is the maximally entangled state $\vert\phi^{+}_d\rangle\langle\phi^{+}_d\vert$, while the corresponding measurements performed by Alice are fixed as
\begin{equation}
\hat{\Pi}^{*a}_{\mu}=\varepsilon_{\eta}(\hat{M}^{a}_{\mu}), \forall a\in\{1,...,d\}, \mu\in\{1,...,N\}.
 \end{equation}
For such a special case, the statement (b) can be expressed in an explicit way: the set of measurements, $\{\varepsilon_{\eta}(\hat{M}^{a}_{\mu})\}$, is incompatible if $\eta>  \mathfrak{g}_{\mathrm{NST}}(\{M^a_{\mu}\},N, d\})$. With the critical visibility defined in Eq.~\eqref{CV}, if $\eta>\eta(\{\hat{M}_{\mu}^{a}\},N,d)$, $\{\varepsilon_{\eta}(\hat{M}_{\mu}^{a})\}$ is incompatible, and if $\eta=\eta(\{\hat{M}_{\mu}^{a}\},N,d)$, $\{\varepsilon_{\eta}(\hat{M}_{\mu}^{a})\}$ is jointly measurable. Thus, one may arrive at the conclusion
\begin{equation}
\eta(\{\hat{M}_{\mu}^{a}\}, N,d)\leq \mathfrak{g}_{\mathrm{NST}}(\{M^a_{\mu}\},N, d).
\end{equation}
For example, let us consider the case where a pair of projective measurements are defined for the qubit case,
\begin{eqnarray}
\label{qubit}
\vert\phi^1_2\rangle&=&\cos\theta\vert\phi^1_1\rangle+\sin\theta\vert\phi^2_1\rangle,\nonumber\\
\vert\phi^2_2\rangle&=-&\sin\theta\vert\phi^1_1\rangle+\cos\theta\vert\phi^2_1\rangle,
\end{eqnarray}
with $\langle\phi^a_{\mu}\vert\phi^b_{\mu}\rangle=\delta_{ab}, a,b, \mu=1,2$. The critical visibility is~\cite{de,car,haa}
\begin{equation}
\label{qucv}
\eta(\{\hat{M}_{\mu}^{a}\})=\frac{1}{\vert\cos\theta\vert+\vert\sin\theta\vert},
\end{equation}
while the geometric NST, whose derivation can be found in the next section, is
\begin{equation}
\label{quNST}
\mathfrak{g}_{\mathrm{NST}}(\{M^a_{\mu}\})=\max\{\vert\cos\theta\vert,\vert\sin\theta\vert\}.
\end{equation}
One may easily verify that the relation $\eta(\{\hat{M}_{\mu}^{a}\})\leq \mathfrak{g}_{\mathrm{NST}}(\{M^a_{\mu}\})$ always holds.

As in Ref.~\cite{bava}, we can address such a problem: which set of projective measurements is optimal for Bob, when the state $W$ is unknown, and $q_{\mu}$ is fixed as $q_{\mu}=1/N$? This problem is equivalent to finding the optimal geometric NST
 \begin{equation}
\mathfrak{g}^*(\{\hat{M}_{\mu}^{a}\}^{\mathrm{opt}}, N,d)=\min_{\{\hat{M}^a_{\mu}\}}\mathfrak{g}_{\mathrm{NST}}(\{\hat{M}_{\mu}\}, N,d),
\end{equation}
with $\{\hat{M}_{\mu}^{a}\}^{\mathrm{opt}}$ as the set of optimal measurements. Except for some special cases, the analytical derivation
of $\mathfrak{g}^*(\{\hat{M}_{\mu}^{a}\}^{\mathrm{opt}}, N,d)$ is usually a difficult task. One of the possible ways to solve this problem is to apply the relation between the optimal geometric NST and the best critical visibility,
\begin{equation}
\eta^*(\{\hat{M}_{\mu}^{a}\}^{\mathrm{\min}}, N,d)\leq \mathfrak{g}^*(\{\hat{M}_{\mu}^{a}\}^{\mathrm{opt}}, N,d),
\end{equation}
whose reason is simple as follows. If the best critical visibility has the property
\begin{equation}
\label{relation}
\eta^*(\{\hat{M}_{\mu}^{a}\}^{\mathrm{\min}}, N,d)= \mathfrak{g}_{\mathrm{NST}}(\{\hat{M}_{\mu}^{a}\}^{\min}, N,d),
\end{equation}
one may immediately conclude that the measurement $\{\hat{M}_{\mu}^{a}\}^{\mathrm{\min}}$ achieving the best critical visibility should be the optimal measurement for attaining $\mathfrak{g}^*$. Under the condition above, the optimal geometric NST is
\begin{equation}
\mathfrak{g}^*( N,d)=\eta^*(\{\hat{M}_{\mu}^{a}\}^{\mathrm{\min}}, N,d).
\end{equation}
Recalling the example in Eq.~\eqref{qubit}, and from Eq.~\eqref{qucv}, one can have $\eta^*=\sqrt{2}/2$ with $\theta=\pi/4$. From Eq.~\eqref{quNST}, one can have the optimal geometric NST $\mathfrak{g}^*=\sqrt{2}/2$ with the same value of $\theta$. The relation in Eq.~\eqref{relation} has been verified for the special case in Eq.~\eqref{qubit}. However, the general condition, under which Eq.~\eqref{relation} always holds, is still unknown.  In the next section, a series of linear steering criteria, which has been obtained in Ref.~\cite{bava}, can be derived from the calculation of the best critical visibility.

\section{optimal linear steering criteria}
\label{Sec4}

\subsection{LHV model for qubit cases}
For qubit cases ($d=2$), a geometric picture is convenient to characterize an arbitrary density matrix $\rho=\frac{1}{2}(I_2+\mathbf{r}\cdot\bm\sigma)$, with $\bm\sigma=(\sigma_x,\sigma_y,\sigma_z)$ the Pauli matrices and a three-dimensional Bloch vector $\mathbf{r}=(r_x,r_y,r_z)$. The geometric length of $\mathbf{r}$ is denoted by $\vert \mathbf{r}\vert=\sqrt{r_x^2+r_y^2+r_z^2}$. Furthermore, the two measurement results by Alice are usually denoted by $a=+,-$. Then, the projectors for the $\mu$-th measurement can be expressed as $\hat{\Pi}^{\pm}_{\mu}=(I_2\pm\mathbf{\hat{ r}}_{\mu}\cdot\bm \sigma)/2$ with $\mathbf{\hat{ r}}_{\mu}$ a unit vector, and the target states can be written as $\hat{\Phi}^{a}_{\mu}=(I_2\pm\mathbf{\hat{ n}}_{\mu}\cdot\bm\sigma)/2$. Based on this assumption, one may introduce a quantity
\begin{equation}
\mathfrak{A }(\mu,\xi)=\mathfrak{p}(+\vert\mu,\xi)-\mathfrak{p}(-\vert\mu,\xi),
\end{equation}
and by the constraints $\mathfrak{p}(+\vert\mu,\xi)+\mathfrak{p}(-\vert\mu,\xi)=1$, it can be obtained that $-1\leq \mathfrak{A }(\mu,\xi)\leq 1.$  In fact, $\mathfrak{A }(\mu,\xi)$ may be viewed as the
pre-determined  value of the operator $\mathbf{\hat{r}}_\mu\cdot\bm\sigma$ in an LHV  model.

A unit vector $\mathbf{\hat{n}}$ can be expressed as $\mathbf{\hat{n}}(\theta,\phi)=\sin\theta\cos\phi \mathbf{\hat{x}}+\sin\theta\sin\phi \mathbf{\hat{y}}+\cos\theta \mathbf{\hat{z}}$, with two parameters $0 \leq\theta\leq\pi,0\leq\phi<2\pi$, and $q(\theta, \phi)$ is the probability density for the measurement $\mathbf{\hat{n}}\cdot\bm\sigma$ performed by Bob, $\int^{\pi}_{0}d\theta\int^{2\pi}_{0}d \phi q(\theta,\phi)=1$. Introducing a vector
\begin{equation}
\mathbf{\bar{r }}=\int^{\pi}_{0}d\theta\int^{2\pi}_{0}d \phi q(\theta,\phi) \mathfrak{A}(\theta,\phi,\xi)\mathbf{\hat{n}}(\theta,\phi),
\end{equation}
the state $\bar{\rho}$ in Eq.~\eqref{rhobar} becomes $\bar{\rho}=(I_2+\mathbf{\bar{\mathbf{r}}}\cdot\bm\sigma)/2$, and, obviously, $\rho_{\times}=(1+\vert\mathbf{\bar{\mathbf{r}}}\vert)/2$. Now, the geometric NST can be reexpressed as
\begin{equation}
\mathfrak{g}=\max_{\{-1\leq \mathfrak{A }(\theta,\phi,\xi)\leq 1\}}\vert \bar{\mathbf{r}}\vert,
\end{equation}
With a suitable basis, $\bar{\mathbf{r}}$ can be along the direction of $\mathbf{\hat{z}}$ and, therefore, $\vert\bar{\mathbf{r}}\vert=\vert \bar{r}_z\vert.$ From the definition of $\bar{\mathbf{r}}$, one can find
$\bar{r}_z=\int_{0}^{\pi}d\theta\int_{0}^{2\pi} d\phi q(\theta,\phi)\mathfrak{A}(\theta,\phi,\xi)\cos\theta$, and,
noting $-1\leq\mathfrak{A}(\theta,\phi,\xi)\leq 1$, one could obtain
\begin{equation}
\label{GNST1}
\mathfrak{g}=\int_{0}^{\pi}d\theta\int_{0}^{2\pi} d\phi q(\theta,\phi)\vert\cos\theta\vert.
\end{equation}
Since it is always possible to find a suitable basis in which $\bar{\mathbf{r}}$ lies along $\hat{\mathbf{z}}$, the above two equations can be viewed as the general formula for calculating the geometric threshold. One may easily verify that~\emph{the geometric threshold, which has been calculated with  $-1\leq \mathfrak{A}(\theta, \phi,\xi)\leq 1$, remains the same if a more restrictive condition $\mathfrak{A}(\theta,\phi,\xi)\in\{-1,+1\}$ is applied.}

Now, assume that the number of the experiment settings $N$ is finite and $q_{\mu}=1/N$ with $\mu\in\{1,2,..,N\}$. The measurements of Bob are denoted by a set of unit vectors $\{\hat{\mathbf{n}}_{\mu}\}$, and there are $2^N$ vectors $\mathbf{\bar{r}}_{\pm\pm...\pm}=\frac{1}{N} \sum_{\mu=1}^N(\pm\mathbf{\hat{n}}_{\mu})$. The geometric NST now takes the form
\begin{equation}
\label{GNST2}
\mathfrak{g}_{\mathrm{NST}}(\{\mathbf{\hat{n}}_{\mu}\}):=\max_{\{\pm\pm...\pm\}}\vert \mathbf{\bar{r}}_{\pm\pm...\pm}\vert.
\end{equation}
With denotations introduced above, the optimal  geometric NST can be reexpressed as
\begin{equation}
\mathfrak{g}^*=\min_{\{\mathbf{\hat{n}}_{\mu}\}}\mathfrak{g}_{\mathrm{NST}}(\{\mathbf{\hat{n}}_{\mu}\}).
\end{equation}

\subsection{Qubit case}
With the  set of unit vectors $\{\mathbf{\hat{n}}_{\mu},\mu=1,2,...N\}$, the geometric averaged fidelity in Eq.~\eqref{GNST} can be rewritten as
\begin{equation}
f_{\mathrm{avg}}=\frac{1}{N}\sum_{\mu=1}^N\langle \mathbf{\hat{r}}_{\mu}\cdot\bm\sigma\otimes\mathbf{\hat{n}}_{\mu}\cdot\bm\sigma\rangle.
\end{equation}
As shown in the end of the above section, there is an indirect way to derive the optimal NST: if $\eta^*(\{\hat{\mathbf{n}}_{\mu}\})= \mathfrak{g}_{\mathrm{NST}}(\{\hat{\mathbf{n}}_{\mu}\})$ can be verified, the set of measurements  $\{\hat{\mathbf{n}}_{\mu}\}$ is optimal for obtaining $\mathfrak{g}^*$. Here, we shall apply this protocol to construct a series of
optimal linear steering criteria in Ref.~\cite{bava}.

(a) For $N=2$ and $3$, the best critical values are $\eta^*(\{\mathbf{\hat{z}},\mathbf{\hat{x}}\})=\sqrt{2}/2$ and  $\eta^*(\{\mathbf{\hat{z}},\mathbf{\hat{x}},\mathbf{\hat{y}}\})=\sqrt{3}/3$, respectively. By simple calculation, one can find $\mathfrak{g}_{\mathrm{NST}}(\{\mathbf{\hat{z}},\mathbf{\hat{x}}\})=\sqrt{2}/2$ and $\mathfrak{g}_{\mathrm{NST}}(\{\mathbf{\hat{z}},\mathbf{\hat{x}},\mathbf{\hat{y}}\})=\sqrt{3}/3$. From these results, two linear steering criteria for steering from Alice to Bob, can be obtained
\begin{eqnarray}
\langle\hat{\mathbf{r}}_1\cdot\bm\sigma\otimes\sigma_x\rangle&+&\langle \hat{\mathbf{r}}_2\cdot\bm\sigma \otimes \sigma_z\rangle> \sqrt{2},\\
\langle \hat{\mathbf{r}}_1\cdot\bm\sigma\otimes \sigma_x\rangle&+&
\langle \hat{\mathbf{r}}_2\cdot\bm\sigma\otimes \sigma_y\rangle+
\langle \hat{\mathbf{r}}_3\cdot\bm\sigma\otimes \sigma_z\rangle> \sqrt{3}.\nonumber
\end{eqnarray}
These two criteria first appeared in Ref.~\cite{can22}, and recently, with analytical derivations, Cavalcanti and Skrzypczyk demonstrated the above inequalities can be returned from the semidefinite program~\cite{Can1}. Here, with the averaged fidelity as the steering parameter, we have proved that these criteria are also optimal.

(b) The planar qubit projective measurements, whose Bloch vectors are confined to the same plane,  have simple experiment implementations~\cite{Joness}. It has been observed that the most incompatible set of $N\in\{2,3,4,5\}$ planar projective qubit measurements is the set of equally spaced measurements \cite{bava}
\begin{equation}
\mathbf{\hat{n}}_{\mu}=\cos\frac{(\mu-1)\pi}{N}\mathbf{\hat{x}}+\sin\frac{(\mu-1)\pi}{N}\mathbf{\hat{y}}.
\end{equation}
with $\mu\in\{1,2,...,N\}.$
The authors also conjectured that this result is valid for any number of planar projective measurements.
From Eq.~\eqref{GNST2}, the geometric NST is
\begin{equation}
\mathfrak{g}_{\mathrm{NST}}(N)=\left\{\begin{array}{l}
    \frac{2}{N}\sum_{m=1}^{N/2}\cos\frac{m\pi}{2N},(N\mathrm{~is~even}),\\ \\
 \frac{2}{N}\sum_{m=0}^{N/2-1}\cos\frac{(2m+1)\pi}{2N}, (N\mathrm{~is~odd}).
  \end{array}\right.
\end{equation}
From $N=2$ to $N=10$, the numerical values of $\mathfrak{g}_{\mathrm{NST}}$  are listed as follows: 0.7071, 0.6667, 0.6533, 0.6472, 0.6440, 0.6420, 0.6407, 0.6399, and 0.6392. These values exactly coincide to the up bound $\eta^*(N, d=2)$. Formally, we have a series of optimal linear criteria for the planar measurements:
\begin{equation}
\frac{1}{N}\sum_{\mu=1}^N\langle\mathbf{\hat{r}}_{\mu}\cdot\bm\sigma\otimes\mathbf{\hat{n}}_{\mu}\cdot\bm\sigma\rangle> \mathfrak{g}_{\mathrm{NST}}(N).
\end{equation}

\subsection{High-dimensional system}
For the two-settings case, the two sets of orthogonal basis, $\{\vert\phi^a_{1}\rangle\}$ and  $\{\vert\phi^b_{2}\rangle\}$ with $a,b\in\{1,2,...d\}$, are supposed to be related to a unitary matrix $U$ with $U_{ab}$ as its matrix elements, $\vert\phi^b_2\rangle=\sum_{a=1}^d U_{ba}\vert\phi^a_1\rangle$. Let the probability for each setting be fixed as $q_1=q_2=1/2$ and, with the deterministic LHV model, one can introduce a series of states $\rho^{ab}$ as $\rho^{ab}=\frac{1}{2}(\hat{\Phi}^{a}_1+\hat{\Phi}^b_2)$ and define the fidelity NST as
 \begin{equation}
 \mathfrak{F}_{\mathrm{NST}}(U)=\max_{a,b\in\{1,2,...d\}}\{\rho^{ab}_{\times}\}.
 \end{equation}
Now, the optimal $\mathrm{\mathrm{NST}}$ can be obtained
\begin{equation}
\mathfrak{F}^{*}=\min_U\mathfrak{F}_{\mathrm{NST}}(U),
\end{equation}
where the optimization is taken over all unitary matrices.

Let us consider a simple case. With $\vert e_1\rangle$ and $\vert e_2\rangle$ a pair of orthogonal  states, a pure state $\vert \phi\rangle$ can be expressed as $\vert\phi\rangle=s\vert e_1\rangle+\sqrt{1-\vert s\vert^2} \vert e_2\rangle$, and for a mixture $\rho=\frac{1}{2}(\vert e_1\rangle\langle e_1\vert+\vert\phi\rangle\langle \phi\vert)$, it can be simply obtained that $\rho_{\times}=\frac{1}{2}(1+\vert s\vert).$ Based on this, there should be $\rho_{\times}^{ab}=(1+\vert U_{ab}\vert)/2$ and
\begin{equation}
\mathfrak{F}_{\mathrm{NST}}(U)=\frac{1}{2}(1+\max_{a,b\in\{1,2,...d\}}\{\vert U_{ab}\vert\}).
\end{equation}
Formally, it is equivalent to $\mathfrak{F}_{\mathrm{NST}}(U)=(1+ \vert U^{\mathrm{opt}}_{ab}\vert)/2$.  In other words, one should select out the optimal unitary element $U^{\mathrm{opt}}_{ab}$, whose mode $\vert U^{\mathrm{opt}}_{ab}\vert$ has the largest value, from all the unitary matrix elements. Consider the constraint for an arbitrary unitary matrix: $\sum_{a=1}^d \vert U_{ab}\vert^2=1$, and the minimum value of $\vert U^{\mathrm{opt}}_{ab}\vert$ should be $1/\sqrt{d}$. The optimal NST now can be obtained:
\begin{equation}
\mathfrak{F}^*=\frac{1}{2}(1+\frac{1}{\sqrt{d}}).
\end{equation}
For the optimal geometric NST, one can have
\begin{equation}
\mathfrak{g}^*=\frac{1}{2}(1+\frac{1}{\sqrt{d}+1}).
\end{equation}
It is known that a set of mutually unbiased bases (MUBs) consists of two or more orthonormal bases $\{\vert\phi_{x}^a\rangle,a=1,2,...,d\}$ in a $d$-dimensional Hilbert space satisfying
\begin{equation}
\vert\langle\phi^a_x\vert\vert\phi^b_y\rangle\vert^2=\frac{1}{d}, \forall a,b\in\{1,...,d\}, x\neq y,
\end{equation}
for all $x$ and $y$~\cite{mubs}, and the MUBs are shown to be the optimal measurements for $\mathfrak{F}^*$. Collecting the above results together, one can have the optimal steering criterion
\begin{equation}
\sum_{a=1}^d\sum_{\mu=1}^2\langle\hat\Pi_{\mu}^a\otimes\hat{\Phi}^{a}_{\mu}\rangle >1+\frac{1}{\sqrt{d}},
\end{equation}
with $\hat\Phi^a_{\mu}$ one of MUBs.
The critical visibility for the two sets of MUBs has already been obtained~\cite{de,car,haa}:
\begin{equation}
\eta(\{\hat\Phi^a_{\mu}\}, N=2,d)=\frac{1}{2}(1+\frac{1}{\sqrt{d}+1}).
\end{equation}

As known in the definition of critical visibility in Eq.~\eqref{CV}, if $\mathfrak{g}^*=\eta(\{\hat\Phi^a_{\mu}\},N,d)$, $\mathfrak{g}^*$ is a tight bound.

\subsection{Some candidates for the optimal criteria}
(a) For $N=4$ and $d=2$, Bavaresco \emph{et. al.} proved that intuitive notions of equally spaced measurements in the Bloch sphere do not correspond to the best measurements~\cite{bava}. For four measurements, there are three coplanar and equally distributed vectors and one vector orthogonal to the former three. With a suitable basis, the four vectors can be expressed as
\begin{eqnarray}
\label{vectors}
\mathbf{\hat{n}}_0&=& \mathbf{\hat{z}},\ \ \mathbf{\hat{n}}_1=-\frac{1}{2}\mathbf{\hat{x}}+\frac{\sqrt{3}}{2}\mathbf{\hat{y}},\nonumber\\
\mathbf{\hat{n}}_2&=& \mathbf{\hat{x}},\ \ \mathbf{\hat{n}}_3=-\frac{1}{2}\mathbf{\hat{x}}-\frac{\sqrt{3}}{2}\mathbf{\hat{y}}.
\end{eqnarray}
From the vectors above, one can choose one of the optimal vectors $\mathbf{\bar{r}}_{\mathrm{+-+-}}=\frac{1}{4}(\mathbf{\hat{n}}_0-\mathbf{\hat{n}}_1+\mathbf{\hat{n}}_2-\mathbf{\hat{n}}_3)$, and the geometric NST is obtained as  $\mathfrak{g}_{\mathrm{NST}}(\{\mathbf{\hat{n}}_{\mu}\})=\vert\mathbf{\bar{r}}_{+-+-}\vert=\sqrt{5}/4\simeq 0.5590$. The best critical visibility is $\eta^*(\{\mathbf{\hat{n}}_{\mu}\})\simeq 0.5544$~\cite{bava}. Since $\eta^*(\{\mathbf{\hat{n}}_{\mu}\})<\mathfrak{g}_{\mathrm{NST}}(\{\mathbf{\hat{n}}_{\mu}\})$, one can not directly prove that the set of measurements in Eq.~\eqref{vectors} are also optimal in obtaining the NST. However, we conjecture that the measurements are optimal for the following two reasons: first, the two values $\eta^*(\{\mathbf{\hat{n}}_{\mu}\})$ and $\mathfrak{g}_{\mathrm{NST}}(\{\mathbf{\hat{n}}_{\mu}\})$ are very close and, second, $\mathfrak{g}_{\mathrm{NST}}(\{\mathbf{\hat{n}}_{\mu}\})=0.5590$,  is better than the one, $\mathfrak{g}_{\mathrm{NST}}=0.5774$, for the case where four vectors are equally distributed in the Bloch sphere. We organize the results above as a steering criterion: if the condition
\begin{equation}
\label{crit5}
\sum_{\mu=0}^3\langle\mathbf{\hat{r}}_{\mu}\cdot\bm\sigma\otimes\mathbf{\hat{n}}_{\mu}\cdot\bm\sigma\rangle>\sqrt{5},
\end{equation}
is satisfied, the state is steerable from Alice to Bob.

(b) Another example is a natural generation of the planar states considered above. A unit vector $\mathbf{\hat{n}}(\phi)=\cos\phi \mathbf{\hat{x}}+\sin\phi \mathbf{\hat{y}}$, is randomly chosen from the $\mathbf{\hat{x}}$-$\mathbf{\hat{y}}$ plane with the probability distribution $q(\phi)=\frac{1}{2\pi}$. Certainly, $\int_{0}^{2\pi}d\phi q(\phi)=1$, and $\mathbf{\bar{r }}$ can be expressed as $\mathbf{\bar{r }}= \int_{0}^{2\pi} d\phi q(\phi)\mathfrak{A}(\phi,\xi){\hat{\mathbf{n}}}(\phi)$. With a suitable basis, the vector $\mathbf{\bar{r }}$ can lie along the direction of $\bar{\mathbf{x}}$, and then, $\mathbf{\bar{r}}=\bar{r}_x\mathbf{\hat{x}}$ with $\bar{r}_y=0$. Therefore, according to the definition of  $\mathbf{\bar{r}}$, one has
\begin{eqnarray}
\vert \mathbf{\bar{r }}\vert&=&\left\vert \int_{0}^{2\pi} q(\phi)\cos\phi\mathfrak{A}(\phi,\xi) d\phi\right\vert\nonumber\\
&\leq&\int_{0}^{2\pi} q(\phi)\vert\cos\phi \vert d\phi=\frac{2}{\pi},
\end{eqnarray}
and the geometric threshold can be obtained as $\mathfrak{g}=2/\pi$. The result can be written as a criterion,
\begin{equation}
\int^{2\pi}_{0}d\phi q(\phi)\langle\bm\sigma\cdot\hat{\mathbf{r}}(\phi)\otimes\bm\sigma\cdot\hat{\mathbf{n}}(\phi)\rangle >\frac{2}{\pi},
\end{equation}
for the state to be  steerable from Alice to Bob.  The result in Ref.~\cite{Joness} has been recovered here.

To show the above threshold is tight, one may use $\phi$ as the local hidden variable, and define $\omega(\phi)=\frac{1}{2\pi}$ for the local states $\{\rho_{\phi}=(I_2+\cos\phi\sigma_x+\sin\phi\sigma_y)/2,\phi\in[0,2\pi)\}$. Consider the case $\tau=I_2/2$ and a depolarizing channel $\varepsilon_{\eta}(\hat{\Psi})=(1-\eta)I_2/2+\eta\hat{\Psi}$. Introducing $\hat{\mathbf{n}}=\cos\alpha \mathbf{\hat{x}}+\sin\alpha \mathbf{\hat{y}}$ for the input states $\hat{\Psi}^{\pm}=(I_2\pm\hat{\mathbf{n}}\cdot\bm\sigma)/2$, and letting $\eta=\pi/2$, the corresponding output could be $\varepsilon_{\eta}(\hat{\Psi}^{\pm})=(I_2\pm\frac{2}{\pi}\hat{\mathbf{n}}\cdot\bm\sigma)/2$. For simplicity, we choose $\alpha=\pi/2$, and $\varepsilon(\hat{\Psi}^{\pm})=(I_2\pm\frac{2}{\pi}\sigma_y)/2$ can  be decomposed with $\{\rho_{\phi}\}$,
\begin{eqnarray}
\frac{1}{2}\varepsilon(\hat{\Psi}^{+})&=&\int_{0}^{2\pi} \mathfrak{p}(+\vert \alpha, \phi)\omega(\phi)\rho(\phi)d\phi,\nonumber\\
\mathfrak{p}(+\vert \alpha, \phi)&=&\left\{\begin{array}{ll}
                                    1,\ 0\leq \phi\leq \pi \\
                                    0,\ \pi\leq \phi\leq 2\pi
                                  \end{array}\right..
                                  \end{eqnarray}
The probability ${p}(+\vert\alpha, \phi)$ can be calculated as $\int_{0}^{2\pi} \mathfrak{p}(+\vert \alpha, \phi)\omega(\phi)d\phi=1/2$. Meanwhile, we have
\begin{eqnarray}
\frac{1}{2}\varepsilon(\hat{\Psi}^{-})&=&\int_{0}^{2\pi} \mathfrak{p}(-\vert \alpha, \phi)\omega(\phi)\rho(\phi)d\phi,\nonumber\\
\mathfrak{p}(-\vert \alpha, \phi)&=&\left\{\begin{array}{ll}
                                    0,\ 0\leq \phi\leq \pi \\
                                    1,\ \pi\leq \phi\leq 2\pi
                                  \end{array}\right..
                                  \end{eqnarray}
Certainly, $ {p}(-\vert\alpha, \phi)=\int_{0}^{2\pi} \mathfrak{p}(-\vert \alpha, \phi)\omega(\phi)d\phi=1/2$.
 The above formula can be easily generalized to the case where $\alpha$ takes an arbitrary value. As a conclusion, we have shown that the assemblage $\{(I_2\pm\frac{2}{\pi}\bm\sigma\cdot\hat{\mathbf{n}})/2\}$ admits a LHS model and  the geometric threshold $\mathfrak{g}_{\mathrm{NST}}=\pi/2$ is a tight bound.

\section{Optimal state-adapted steering criteria}
\label{Sec5}
 
\subsection{CHSH-like inequality}
In Ref.~\cite{Can3}, the authors derived an EPR-steering analog of the CHSH inequality, and it is an inequality that is necessary and sufficient to demonstrate EPR steering in a scenario involving only correlations between two dichotomic measurements on each subsystem.  However, this inequality requires that the measurements by the trusted party are mutually unbiased. Recently, Girdhar and Cavalcanti produced a necessary and sufficient steering inequality~\cite{Girdhar} in the same CHSH scenario to that in Ref.~\cite{Can3}, and it is applicable to any pair of projective measurements at the trusted part. This CHSH-like inequality has a very interesting property: \emph{all two-qubit states that are steerable via this inequality are Bell nonlocal.}

Considering the fact that the probabilities for the experiment settings are controllable parameters for Alice (or Bob), we shall provide a simple criterion, a sufficient condition for steerability of a two-qubit state from Alice to Bob, which also has the property discussed above.

At first, let us consider the case where Bob's measurements are MUBs: $\mathbf{\hat{n}}\cdot\bm\sigma$ and $\mathbf{\hat{n}}_{\bot}\cdot\bm\sigma$, where  $\mathbf{\hat{n}}$ and $\mathbf{\hat{n}}_{\bot}$ are orthogonal to each other.  With $q(\mathbf{\hat{n}})$
and $q(\mathbf{\hat{n}}_{\bot})$ the probability distributions for each measurement, respectively, we can calculate the geometric NST,
$\mathfrak{g}_{\mathrm{NST}}=\sqrt{q^2(\mathbf{\hat{n}})+q^2(\mathbf{\hat{n}}_{\bot})}$, and the geometric averaged fidelity,
 $f_{\mathrm{avg}}=q(\mathbf{\hat{n}})\langle\mathbf{\hat{a}}\otimes \mathbf{\hat{n}}\rangle +q(\mathbf{\hat{n}}_{\bot})\langle \mathbf{\hat{b}}\otimes \mathbf{\hat{n}}\rangle$, with the denotation $\langle\mathbf{\hat{a}}\otimes \mathbf{\hat{n}}\rangle \equiv\langle
\mathbf{\hat{a}}\cdot\bm{\sigma}\otimes\mathbf{\hat{n}}\cdot\bm{\sigma}\rangle$.
Now, we define $\mathfrak{R}_2$ as the ration of the geometric averaged fidelity and the geometric NST, $\mathfrak{R}_2:=f_{\mathrm{avg}}/\mathfrak{g}_{\mathrm{NST}}$, and a criterion for steering from Alice to Bob can be obtained
\begin{equation}
\label{crir2}
\mathfrak{R}_2:=\frac{q(\mathbf{\hat{n}})\langle \hat{\mathbf{a}}\otimes \mathbf{\hat{n}}\rangle}{\sqrt{q^2(\mathbf{\hat{n}})+q^2(\mathbf{\hat{n}}_{\bot})}}
+\frac{q(\mathbf{\hat{n}}_{\bot})\langle \hat{\mathbf{b}}\otimes \mathbf{\hat{n}}_{\bot}\rangle}{\sqrt{q^2(\mathbf{\hat{n}})+q^2(\mathbf{\hat{n}}_{\bot})}}>1.
\end{equation}
From the CHSH inequality~\cite{chsh}, one may have a sufficient condition for a state to be Bell-nonlocal
\begin{equation}
\label{cribell}
\langle \mathbf{\hat{a}}\otimes (\mathbf{\hat{n}}_1-\mathbf{\hat{n}}_2)\rangle +\langle \mathbf{\hat{b}}\otimes (\mathbf{\hat{\mathbf{\hat{n}}}}_1+\mathbf{\hat{n}}_2)\rangle>2.
\end{equation}
As in Refs.~\cite{pop,h3},  one can introduce a pair of orthogonal unit vectors $\mathbf{\hat{n}}'$ and $\mathbf{\hat{n}}_{\bot}'$, and the unit vectors ${\mathbf{\hat{n}}}_i$ can be expanded as
\begin{equation}
\mathbf{\hat{n}}_1=\cos\theta\mathbf{\hat{n}}'+\sin\theta \mathbf{\hat{n}}_{\bot}',\ \mathbf{\hat{n}}_2=-\cos\theta\mathbf{\hat{n}}'+\sin\theta \mathbf{\hat{n}}_{\bot}',
\end{equation}
or equivalently
\begin{equation}
\label{vectors2}
\mathbf{\hat{n}}_1-\mathbf{\hat{n}}_2=2\cos\theta \mathbf{\hat{n}}',\ \mathbf{\hat{n}}_1+\mathbf{\hat{n}}_2=2\sin\theta \mathbf{\hat{n}}'_{\bot}.
\end{equation}
Now, we introduce another pair of orthogonal unit vectors $\mathbf{\hat{n}}, \mathbf{\hat{n}}_{\bot}$,
\begin{equation}
\mathbf{\hat{n}}=\frac{\cos\theta}{\vert\cos\theta\vert}\mathbf{\hat{n}}',\ \mathbf{\hat{n}}_{\bot}=\frac{\sin\theta}{\vert\sin\theta\vert} \mathbf{\hat{n}}'_{\bot},
\end{equation}
and Eq.~\eqref{vectors2} becomes
\begin{equation}
\mathbf{\hat{n}}_1-\mathbf{\hat{n}}_2=2\vert\cos\theta\vert \mathbf{\hat{n}},\ \mathbf{\hat{n}}_1+\mathbf{\hat{n}}_2=2\vert\sin\theta\vert \mathbf{\hat{n}}_{\bot}.
\end{equation}
Putting it back into Eq.~\eqref{cribell}, the criterion for the Bell-nonlocal state comes to
\begin{equation}
\vert\cos\theta\vert\langle \mathbf{\hat{a}}\otimes \mathbf{\hat{n}}\rangle  +\vert\sin\theta\vert \langle \hat{\mathbf{b}} \otimes \mathbf{\hat{n}}_{\bot}\rangle>1.
\end{equation}
Comparing it with the steering criterion in Eq.~\eqref{crir2}, it could be found that the two criteria are very similar. To avoid the misunderstanding of the similarity, two operators could be introduced
\begin{eqnarray}
\hat{T}_{\mathrm{\mathrm{steer}}}&=&\frac{q(\mathbf{\hat{n}})\hat{\mathbf{a}}\cdot\bm\sigma\otimes\hat{\mathbf{n}}\cdot\bm\sigma
+q(\mathbf{\hat{n}}_{\bot})\hat{\mathbf{b}}\cdot\bm\sigma\otimes\mathbf{\hat{n}}_{\bot}\cdot\bm\sigma}
 {\sqrt{q^2(\mathbf{\hat{n}})+q^2(\mathbf{\hat{n}}_{\bot})}},\nonumber\\
 \hat{T}_{\mathrm{CHSH}}&=&\vert \cos\theta\vert\hat{\mathbf{a}}\cdot\bm\sigma\otimes\hat{\mathbf{n}}\cdot\bm\sigma+\vert \sin\theta\vert \hat{\mathbf{b}}\cdot\bm\sigma\otimes\mathbf{\hat{n}}_{\bot}\cdot\bm\sigma,\nonumber
 \end{eqnarray}
 and it can be shown that they are equal to each other: $\hat{T}_{\mathrm{\mathrm{steer}}}=\hat{T}_{\mathrm{CHSH}}$ under the one-to-one mapping,
\begin{equation}
\vert \cos\theta\vert=\frac{q(\mathbf{\hat{n}})}{\sqrt{q^2(\mathbf{\hat{n}})+q^2(\mathbf{\hat{n}}_{\bot})}},\ \vert\sin\theta\vert=\frac{q(\mathbf{\hat{n}}_{\bot})}{\sqrt{q^2(\mathbf{\hat{n}})+q^2(\mathbf{\hat{n}}_{\bot})}}.\nonumber
\end{equation}
Based on the results above, one may conclude that if a state has been verified to be steerable from Alice to Bob by the criterion in Eq.~\eqref{crir2}, it must be Bell nonlocal.

\subsection{The case with known two-qubit state}
In the previous  section, we have considered the case where the state is unknown. In the following, we will address another problem: when the state $W$ and Bob's measurements are known, which kinds of criteria are optimal for steering from Alice to Bob? At first, we introduce $\langle \mathbf{\hat{a}}\otimes \mathbf{\hat{n}}\rangle_{\max}\equiv\max_{\{\mathbf{\hat{a}}\}} \mathrm{Tr} [W\mathbf{\hat{a}}\cdot\bm\sigma\otimes\mathbf{\hat{n}}\cdot\bm\sigma]$ for the best correlation when the two-qubit state  and measurement of Bob are fixed by $\hat{\mathbf{n}}$. Let $\{\mathbf{\hat{n}} , \mathbf{\hat{n}}_{\bot}, \mathbf{\hat{t}}\}$ be a set of mutually orthogonal unit vectors. Another set of mutually orthogonal unite vectors $\{\mathbf{\hat{x}}, \mathbf{\hat{y}}, \mathbf{\hat{z}}\}$, which are state-dependent, is defined as follows: $\langle \mathbf{\hat{a}}\otimes \mathbf{\hat{z}}\rangle_{\max}= \max_{\{\mathbf{\hat{n}}\}}\langle \mathbf{\hat{a}}\otimes \mathbf{\hat{n}}\rangle_{\max}$, $\langle \mathbf{\hat{a}}\otimes \mathbf{\hat{x}}\rangle_{\max}= \max_{\mathbf{\{\hat{n}}_{\bot}\}}\langle \mathbf{\hat{a}}\otimes \mathbf{\hat{n}}_{\bot}\rangle_{\max}$,
and $\langle \mathbf{\hat{a}}\otimes \mathbf{\hat{y}}\rangle_{\max}= \max_{\{\mathbf{\hat{t}}\}}\langle \mathbf{\hat{a}}\otimes \mathbf{\hat{t}}\rangle_{\max}$, where the three optimalities are taken under the constraint that $\{\mathbf{\hat{n}}, \mathbf{\hat{n}}_{\bot}, \mathbf{\hat{t}}\}$ is an orthogonal basis. In general, we suppose $\langle \mathbf{\hat{a}}\otimes \mathbf{\hat{z}}\rangle_{\max}\geq \langle \mathbf{\hat{a}}\otimes \mathbf{\hat{x}}\rangle_{\max}\geq \langle \mathbf{\hat{a}}\otimes \mathbf{\hat{y}}\rangle_{\max}.$ With the set of unambiguously defined vectors $\{\mathbf{\hat{x}}, \mathbf{\hat{y}}, \mathbf{\hat{z}}\}$, Eq.~\eqref{crir2} can be rewritten in a more compact way:
\begin{equation}
\mathfrak{R}_2:=\frac{q_z\langle \mathbf{\hat{a}}\otimes \mathbf{\hat{z}}\rangle}{\sqrt{q^2_x+q^2_z}}+\frac {q_x\langle \mathbf{\hat{b}}\otimes \mathbf{\hat{x}}\rangle}{\sqrt{q^2_x+q^2_z}}>1.
\end{equation}
To make $\mathfrak{R}_2$ have a greater value, the vectors $\mathbf{\hat{a}}$ and $\mathbf{\hat{b}}$ should be fixed by requiring that the correlations $\langle \mathbf{\hat{a}}\otimes \mathbf{\hat{z}}\rangle$ and $\langle \mathbf{\hat{b}}\otimes \mathbf{\hat{x}}\rangle$ attain the maximum values, respectively. The procedure for fixing these vectors can be shown via an explicit example, and one can consider an arbitrary pure state
\begin{equation}
\label{pure}
|\psi\rangle=\cos\frac{\gamma}{2}\vert 00\rangle+\sin\frac{\gamma}{2}\vert 11\rangle,
\end{equation}
with $0\leq \gamma\leq\pi/2$. Letting $\mathbf{\hat{a}}=\cos\theta_1\mathbf{\hat{z}}+\sin\theta_1\cos\phi_1\mathbf{\hat{x}}+\sin\theta_1\sin\phi_1\mathbf{\hat{y}}$ and $\mathbf{\hat{n}}=\cos\theta\mathbf{\hat{z}}+\sin\theta\cos\phi\mathbf{\hat{x}}+\sin\theta\sin\phi\mathbf{\hat{y}}$, we have
$\langle \mathbf{\hat{a}}\otimes \mathbf{\hat{n}}\rangle=\cos\theta_1\cos\theta+\sin\gamma\sin\theta_1\sin\theta\cos(\phi_1+\phi)$, and obtain
\begin{equation}
\langle \mathbf{\hat{a}}\otimes \mathbf{\hat{n}}\rangle_{\max}=\sqrt{\cos^2\theta+\sin^2\gamma\sin^2\theta},
\end{equation}
by requiring $\tan \theta_1=\sin\gamma\tan \theta$, and $\phi_1=-\phi$. From it, there should be
\begin{equation}
\langle \hat{\mathbf{a}}\otimes \mathbf{\hat{z}}\rangle_{\max}=1, \langle \mathbf{\hat{b}}\otimes \mathbf{\hat{x}}\rangle_{\max}=\sin\gamma.
\end{equation}
Via the Cauchy-Schwarz inequality, the optimal choice of $q_z$ and $q_x$ is 
\begin{eqnarray}
q_z&=&\frac{\langle \hat{\mathbf{a}}\otimes \mathbf{\hat{z}}\rangle_{\max}}{\langle \hat{\mathbf{a}}\otimes \mathbf{\hat{z}}\rangle_{\max}+
\langle \mathbf{\hat{b}}\otimes \mathbf{\hat{x}}\rangle_{\max}},\\
q_x&=&\frac{\langle \hat{\mathbf{b}}\otimes \mathbf{\hat{x}}\rangle_{\max}}{\langle \hat{\mathbf{a}}\otimes \mathbf{\hat{z}}\rangle_{\max}+
\langle \mathbf{\hat{b}}\otimes \mathbf{\hat{x}}\rangle_{\max}},
\end{eqnarray}
respectively. By collecting the above results together, one has an optimal steering criterion
\begin{equation}
\label{crit3}
\mathfrak{R}^{\max}_2=\sqrt{\langle \hat{\mathbf{a}}\otimes \mathbf{\hat{z}}\rangle_{\max}^2+
\langle \mathbf{\hat{b}}\otimes \mathbf{\hat{x}}\rangle_{\max}^2}>1,
\end{equation}
which is obviously state-dependent. For the pure state $|\psi\rangle$ in Eq~\eqref{pure}, $\mathfrak{R}^{\max}_2= \sqrt{1+\sin^2\gamma}$, and one may conclude that all entangled pure states ($\sin\gamma\neq 0$) are steerable from Alice to Bob. For the Werner state in Eq.~\eqref{isotopic} when $d=2$, $\mathfrak{R}^{\max}_2=\sqrt{2}\eta$, and the steerable condition is $\eta>\sqrt{2}/2$.

As a simple generation of the above criterion, we consider the case where Bob's measurements are fixed as $\{\sigma_x,\sigma_y, \sigma_z\}$.
The geometric NST now is $\mathfrak{g}_{\mathrm{NST}}=\sqrt{q_x^2+q_y^2+q_z^2}$, while  the geometric averaged fidelity is $f_{\mathrm{avg}}=q_z\langle \mathbf{\hat{a}}\otimes \mathbf{\hat{z}}\rangle+q_x
\langle \mathbf{\hat{b}}\otimes \mathbf{\hat{x}}\rangle +q_y\langle \mathbf{\hat{c}}\otimes \mathbf{\hat{y}}\rangle.$ The
ration $\mathfrak{R}_{3}:=f_{\mathrm{avg}}/\mathfrak{g}_{\mathrm{NST}}$, can be calculated as
\begin{equation}
\mathfrak{R}_3=\frac{q_z\langle \mathbf{\hat{a}}\otimes \mathbf{\hat{z}}\rangle+q_x
\langle \mathbf{\hat{b}}\otimes \mathbf{\hat{x}}\rangle +q_y\langle \mathbf{\hat{c}}\otimes \mathbf{\hat{y}}\rangle}{\sqrt{q_x^2+q_y^2+q_z^2}}.
\end{equation}
In a similar way with the two setting case, by fixing the probabilities as
\begin{eqnarray}
q_z&=&\frac{\langle \mathbf{\hat{a}}\otimes \mathbf{\hat{z}}\rangle_{\max}}{\langle \mathbf{\hat{a}}\otimes \mathbf{\hat{z}}\rangle_{\max}
+\langle \mathbf{\hat{b}}\otimes \mathbf{\hat{x}}\rangle_{\max}+\langle \mathbf{\hat{c}}\otimes \mathbf{\hat{y}}\rangle_{\max} },\nonumber\\
q_x&=&\frac{\langle \mathbf{\hat{b}}\otimes \mathbf{\hat{x}}\rangle_{\max}}{\langle \mathbf{\hat{a}}\otimes \mathbf{\hat{z}}\rangle_{\max}
+\langle \mathbf{\hat{b}}\otimes \mathbf{\hat{x}}\rangle_{\max}+\langle \mathbf{\hat{c}}\otimes \mathbf{\hat{y}}\rangle_{\max} },\\
q_y&=&\frac{\langle \mathbf{\hat{c}}\otimes \mathbf{\hat{y}}\rangle_{\max}}{\langle \mathbf{\hat{a}}\otimes \mathbf{\hat{z}}\rangle_{\max}
+\langle \mathbf{\hat{b}}\otimes \mathbf{\hat{x}}\rangle_{\max}+\langle \mathbf{\hat{c}}\otimes \mathbf{\hat{y}}\rangle_{\max} },\nonumber
\end{eqnarray}
an optimal steering criterion can be obtained:
\begin{equation}
\label{crit4}
\sqrt{\langle \mathbf{\hat{a}}\otimes \mathbf{\hat{z}}\rangle^2_{\max}
+\langle \mathbf{\hat{b}}\otimes \mathbf{\hat{x}}\rangle^2_{\max}+\langle \mathbf{\hat{c}}\otimes \mathbf{\hat{y}}\rangle^2_{\max} }>1.
\end{equation}
For the pure states, $\mathfrak{R}^{\max}_3=\sqrt{1+2\sin^2\gamma}$, all pure entangled states are steerable from Alice to Bob according to the criterion above. Meanwhile, $\mathfrak{R}^{\max}_3=\sqrt{3}\eta$ holds for the Werner state in Eq.~\eqref{isotopic} when $d=2$, and the criterion, $\eta>\sqrt{3}/3$, is better than that for two measurement settings. Besides the theoretical application in judging whether a state is steerable from Alice to Bob, the argument in the derivation of these two criteria in Eq.~\eqref{crit3} and Eq.~\eqref{crit4} is also applicable for designing an optimal experiment protocol if the state shared by Alice and Bob has already been known.

In the end of this section, we propose a criterion for the case where the experiment settings take a
continuous form. Formally, the experiment setting of Bob is denoted by $\mathbf{\hat{n}}\cdot\bm \sigma$ with $\mathbf{\hat{n}}$ an arbitrary unit vector in the Bloch sphere. As in the derivations of $\mathfrak{R}_2$ and $\mathfrak{R}_3$, we fix the probability distribution as
\begin{equation}
q(\mathbf{\hat{n}})=\frac{\mathbf{\langle\hat{a}}\otimes \mathbf{\hat{n}}\rangle_{\max}}{\int\int d\tau \langle \mathbf{\hat{a}}\times \mathbf{\hat{n}}\rangle_{\max}},
\end{equation}
with $ d\tau= \frac{1}{4\pi}\sin\theta d\theta d\phi,\ 0\leq\theta\leq \pi$ and $0 \leq \phi<2\pi$. The geometric averaged fidelity
can be calculated as
\begin{equation}
f_{\mathrm{avg}}= \int\int d\tau q(\mathbf{\hat{n}})\mathbf{\langle\hat{a}}\otimes \mathbf{\hat{n}}\rangle_{\max},
\end{equation}
with  $\mathbf{\bar{r}}=\int\int d\tau q(\mathbf{\hat{n}})\mathfrak{A}(\mathbf{\hat{n}},\xi)\mathbf{\hat{n}}$, and the geometric NST
is $\mathfrak{g}_{\mathrm{NST}}=\max_{\{\mathfrak{A}(\mathbf{\hat{n}},\xi)\in(1,-1)\}}\vert \mathbf{\bar{r}}\vert$. Formally, one can introduce a criterion as follows
\begin{equation}
\label{crit6}
\mathfrak{R}_{\infty}:=\frac{f_{\mathrm{avg}}}{\mathfrak{g}_{\mathrm{NST}}}>1.
\end{equation}
For the Werner state in Eq.~\eqref{isotopic} when $d=2$, $\mathbf{\langle\hat{a}}\otimes \mathbf{\hat{n}}\rangle_{\max}=\eta$,  $q(\mathbf{\hat{n}})=1$ and
$f_{\mathrm{avg}}=\eta$. Using Eq.~\eqref{GNST1}, one can come to $\mathfrak{g}_{\mathrm{NST}}=1/2$ and  $\mathfrak{R}_{\infty}=2\eta$. The criterion, $\eta>1/2$, is obtained for the steerable Werner state.

\section{Conclusions}
In conclusion, the averaged fidelity is taken as the steering parameter, and a general scheme has been developed for designing the optimal steering criteria. Two kinds of optimal steering criteria, which are based on the conditions of whether the bipartite is known or not, have been constructed.

With the fact that the relation between joint measurability and steering has been discussed only in the recent six years~\cite{quint,ula,UULA,Kiukas}, it is easy to see that the basic properties of the linear steering inequality, violation of which means that the measurements of Alice are incompatible, have seldom been discussed before. In the present work, it has been shown that the two conditions---(a) the state is steerable from Alice to Bob and (b) the measurements performed by Alice must be incompatible---are necessary  for the linear steering inequality to be violated.  For the set of measurements performed by Bob, besides the critical visibility \cite{bava}, the NST can be introduced to quantify its ability to detect steering. The relation between critical visibility and NST can be applied to design the optimal steering criteria.

From the experimental viewpoint, the threshold of the criterion in Eq.~\eqref{crit5} is less than ones of  the two-setting and three-setting linear inequalities proposed in Ref.~\cite{can22}. We hope this can be applied in experiments to verify whether an unknown two-qubit state is steerable from Alice to Bob.  From the theoretical view, beyond $T$ states~\cite{jev, ngu}, analytic necessary and sufficient conditions for two-qubit steerability are still not known. In the present work, using the criteria in Eq.~\eqref{crit6}, one can have a steering criterion, $\eta>1/2$,  for the steerable Werner state. This criterion is optimal since we know that the Werner state is unsteerable if $\eta=1/2$~\cite{Wiseman1}.

Whether our scheme can offer optimal criteria for other types of states will be discussed in our  future works, and we expect that the results in this work could lead to further theoretical or experimental consequences.

\acknowledgements
This work was supported by the National Natural Science Foundation of China under Grants No.~11405136 and No.~11947404, and the Fundamental Research Funds for the Central Universities under Grant No.~2682019LK11.

\end{document}